# Flux pinning and phase separation in oxygen rich $La_{2-x}Sr_xCuO_{4+y}$ system.

Hashini E. Mohottala[1], B. O. Wells[1], J. I. Budnick[1], W. A. Hines[1], Ch. Niedermayer[2]

F. C. Chou[3]

[1] *Department of Physics University of Connecticut, Storrs, CT 06269, USA*
[2] *Laboratory for Neutron Scattering, ETHZ & PSI, CH-5232 Villigen PSI, Switzerland*
[3] *Center for Condensed Matter Sciences, National Taiwan University, Taipei 10617, Taiwan*

**Abstract**

We have studied the magnetic characteristics of a series of super-oxygenated $La_{2-x}Sr_xCuO_{4+y}$ samples. As shown in previous work, these samples spontaneously phase separate into an oxygen rich superconducting phase with a $T_C$ near 40 K and an oxygen poor magnetic phase that also orders near 40 K. All samples studied are highly magnetically reversible even to low temperatures. Although the internal magnetic regions of these samples might be expected to act as pinning sites, our present study shows that they do not favor flux pinning. Flux pinning requires a matching condition between the defect and the superconducting coherence length. Thus, our results imply that the magnetic regions are too large to act as pinning centers. This also implies that the much greater flux pinning in typical $La_{2-x}Sr_xCuO_4$ materials is the result of nanoscale inhomogeneities that grow to become the large magnetic regions in the super-oxygenated materials. The superconducting regions of the phase separated materials are in that sense cleaner and more homogenous than in the typical cuprate superconductor.

## I. Introduction

$La_2CuO_4$ holds an important place in studies of high temperature superconductors (HTSC). It is the parent compound of the first cuprate superconductor discovered by Bednorz and Mueller [1]. Amongst the most intensively studied cuprate compounds, doped $La_2CuO_4$ has the simplest structure. The system can be hole doped to produce a superconductor either by cation substitution ($La_{2-x}M_xCuO_4$) or intercalation of excess oxygen ($La_2CuO_{4+y}$), or both [2-5]. The standard phase diagram for the cuprates is based upon $La_{2-x}Sr_xCuO_4$ [6]. It is also in doped versions of $La_2CuO_4$ that the well known $1/8^{th}$ anomaly occurs: superconductivity is at least partially suppressed and an incommensurate ordered magnet phase appears, often described as a stripe phase [7]. Previous studies have shown that doped holes in the system do not necessarily have a uniform density but may phase separate into hole rich phase and hole poor phase. There is evidence for such phases existing with a disordered, very short range nature in cation doped $La_2CuO_4$ but may form much larger domains in systems doped via interstitial oxygen [8-10]. Our own recent studies on co-doped $La_{2-x}Sr_xCuO_{4+y}$ identified the coexistence of separate phases consisting of a $T_C$ = 40 K optimally doped superconductor ($n_h$ = 0.16) and a spin density wave, or stripe phase, that also orders near 40 K ($n_h$ = $1/8^{th}$) [10]. These phases occur at a given hole density regardless of the specific amount of Sr (x) or O (y), thus indicating that the phase separation is driven by the physics of the doped holes themselves.

Much research has focused on the complex interplay between the Cu moments and superconducting properties of cuprate superconductors [11-13]. One consequence of this interplay is the field dependent magnetic response of the superconducting properties, such as hysteretic diamagnetic response, flux trapping, and irreversibility. A primary motivation for such studies is the desire to improve the current carrying capabilities of superconducting cables by increasing flux pinning. For the purest materials that can be made, such pinning is generally rather large in the cuprates compared to traditional, elemental superconductors. Although it is not a primary candidate for cables, some such experiments have been performed on $La_{2-x}Sr_xCuO_4$ [14, 15]. Very few such studies have been reported for super-oxygenated $La_{2-x}Sr_xCuO_4$. In this study we report the magnetic response of the superconducting state of several samples of super-oxygenated $La_{2-x}Sr_xCuO_{4+y}$ in large fields. Other studies have shown that the trapped flux and thus the critical current density can be increased by introducing artificial pinning sites through the use of either substitutional defects or radiation damage [16-19]. Since the super-oxygenated compounds have non-superconducting regions it might be expected that these would act as pinning centers as well. However, we have found that in fact flux pinning in the super-oxygenated materials is substantially reduced compared to purely cation doped $La_{2-x}Sr_xCuO_4$. We argue that this is evidence that the typical cation doped lanthanum cuprate may contain nanoscale inhomogeneities that are effective in trapping flux while in the super-oxygenated material such separate domains grow to a length scale on the order of microns and are thus ineffective at pinning flux.

The layout of this paper is as follows. Under section **II** we discuss the sample preparation techniques. Magnetization experiments come under section **III** and there we discuss both temperature dependence and the field dependence of magnetization measurements. We will discuss and compare our data with the literature in section **IV** and in **V** we will conclude our work.

**II. Sample Preparation**

In this work, we investigated a series of $La_{2-x}Sr_xCuO_4$ samples by varying the Sr contents (x= 0, 0.04 and 0.09). The samples consisted of both crucible grown and float zone grown single crystals. $La_2CuO_4$ (x = 0) was prepared by slow cooling from the melt in a Pt crucible [20]. And the remaining two samples were traveling solvent float-zone-grown single crystals and were synthesized as discussed in ref. [21]. Initially, neither the x = 0 nor the x = 0.04 sample was superconducting, the former was antiferromagnetic and the latter apparently was in the intermediate spin glass regime [22]. The other sample was superconducting with a $T_C$ that is consistent with the well-known phase diagram for $La_{2-x}Sr_xCuO_4$ [23]. We intercalated excess oxygen into all of our samples using wet chemical techniques as discussed in ref. 3. The process took a considerable time (in some cases it took months) to produce well oxidized homogeneous samples.

**III. Magnetization Experiments**

In order to obtain information concerning the superconducting transition and flux penetration in the superconducting state, a variety of magnetization experiments were carried out. After intercalating excess oxygen into the samples, the magnetization was measured using a Quantum Design MPMS SQUID magnetometer. Magnetization vs. temperature scans were obtained at both zero field cooling (ZFC) and field cooling (FC) conditions with a small external magnetic field (~ 10 Oe) along the c axis of the single crystals As shown in Fig. 1, after oxidation, all samples developed a Meissner response (FC response) with a superconducting transition temperature near $T_C$ = 40 K. The superconducting phase is similar for all samples, but the superconducting volume fractions differ from sample to sample. According to our previous μSR studies the sample with x = 0 shows the highest magnetic volume fraction and the one with x = 0.04 shows the lowest magnetic volume fraction [10]. The results reported here are consistent with the previous work.

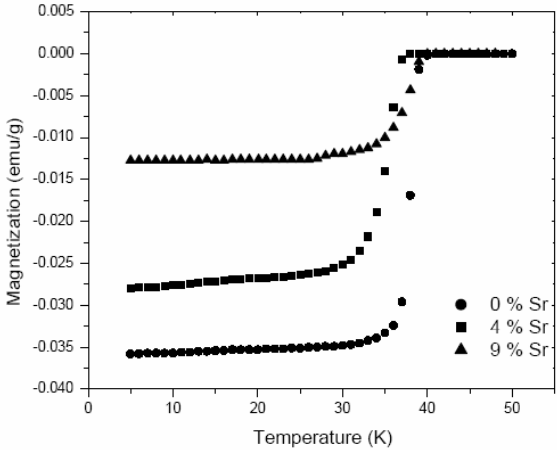

*FIG. 1* FC magnetization data obtained at 10 Oe for a series of super-oxygenated $La_{2-x}Sr_xCuO_{4+y}$ samples. The Meissner fraction varies from sample to sample. However the superconducting transition temperature is nearly the same in all samples (i.e. $T_C = 40$ K).

The key measurements taken for this work are the magnetization versus applied field scans at a variety of temperatures in the superconducting state. These experiments were carried out in order to investigate the flux pinning in our samples. The measurements were performed in the same SQUID magnetometer. For these measurements, the samples were aligned with the magnetic field either parallel or perpendicular to the crystalline c axis. The samples were ZF cooled to the desired temperature and then the field was swept to 50 kOe and back to 0 Oe. Occasionally we performed a full hysteresis loop by sweeping the applied field to 50 kOe, then -50 kOe, and finally back to zero.

Typical magnetization results for our samples are reported in Fig. 2 and Fig. 3. In Fig. 2 the magnetization curves are shown for the case where the magnetic field is applied parallel to the c axis for the samples with Sr content x = 0, 0.04 and 0.09. In Fig. 3 similar results are shown for the case where the field is applied perpendicular to the c axis for the samples with x = 0, 0.04 and 0.09. These curves can be characterized by two parameters. The first is the openness of the curves, $\Delta M = M(H+) - M(H-)$, with $M(H+)$ the value of M at a given H upon increasing the field and $M(H-)$ the value of M at a given H upon decreasing the field in the hysteresis loop. The value of $\Delta M$ is determined by the amount of flux trapped in the sample upon reversal of the applied field and, according to Bean's critical model calculations, it is related to the critical current density, $J_C$ [24]. The absolute value of $J_C$ depends on the sample type and the shape; however in general $J_C$ is directly proportional to $\Delta M$. As can be seen in Fig. 2, at 10 K the three samples have an average $\Delta M_{max}$ of about 15 emu/g. As shown in Fig. 3 when the field is applied perpendicular to the crystallographic c axis the average $\Delta M_{max}$ reduces to about 10 emu/g. These values are about an order of magnitude smaller than the previously reported values for other high temperature superconductors. Since $J_C$ is directly proportional to $\Delta M$, we can conclude that the current density for our samples is about an order of magnitude smaller than that of the reported values for other high $T_C$ materials [14, 25-26].

The second parameter derived from the magnetization data is the irreversibility field, $H_{irr}$, which is the field at which the magnetization upon decreasing the applied field deviates from the magnetization upon increasing the field. This number is related to the pinning energy for flux vortices. The sample to sample variation we observe for $H_{irr}$ is somewhat larger than that for $\Delta M_{max}$. Any of these variations are not obviously correlated with materials parameters such as Sr content, so we assume they are related to overall crystal defect levels. However, this variation is small compared to the results from similar high temperature superconductors with other dopant ions [27]. Thus we can make meaningful comparisons of the average values we find for $H_{irr}$ with other high temperature superconductor results from the literature. At 10 K, our samples have an average value of $H_{irr}$ near 18 kOe. From Fig. 3, it can be seen that there is a substantially larger $H_{irr}$ when the field is applied perpendicular to the c-axis, although the total amount of trapped flux, $\Delta M_{max}$ is not larger. The average number for $H_{irr}$ in this case at T = 10 K is near 50 kOe. Our results, most notable strongly reversible magnetic response, are in good agreement with a previously reported study of a crucible grown, super-oxygenated crystal of $La_2CuO_{4+y}$ that was most likely phase separated in the same manner as our samples [27].

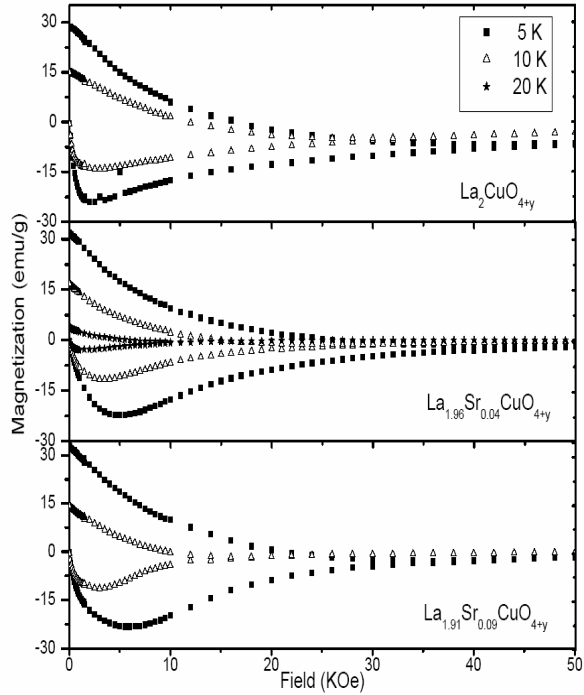

*FIG. 2. Dependence of magnetization on magnetic field (increasing and then decreasing) for various temperatures. The applied magnetic field is parallel to the c axis. All the samples show the reversibility of magnetization at large fields. With the increase of temperature the reversibility increases.*

### IV. Discussion

The most useful published data for comparison to our flux pinning data is a similar single crystal study of $La_{2-x}Sr_xCuO_4$ performed by Kodama et al. [14]. In Fig. 1 of Kodama et al., similar magnetization versus field scans are shown for several single crystals, all with applied magnetic field parallel to the c axis. The most relevant samples for comparison is that for which x = 0.14 and 0.15, roughly similar in total hole concentration as our samples. At 10 K, Kodama et al. find an irreversibility field, $H_{irr}$, well above 50 kOe and a maximum opening $\Delta M$ of about 140 emu/g. Fig. 4 shows the $\Delta M$ values derived from Fig. 1 in Kodama et al and the $\Delta M$ values calculated at 10 K for our samples. The super-oxygenated samples have far less flux pinning. The $\Delta M$ values for optimally doped $La_{2-x}Sr_xCuO_4$ are approximately an order of magnitude larger than those we observe in the super-oxygenated samples. $H_{irr}$ is not well determined for the $La_{2-x}Sr_xCuO_4$ samples but appears to be five times or greater than for the super-oxygenated samples.

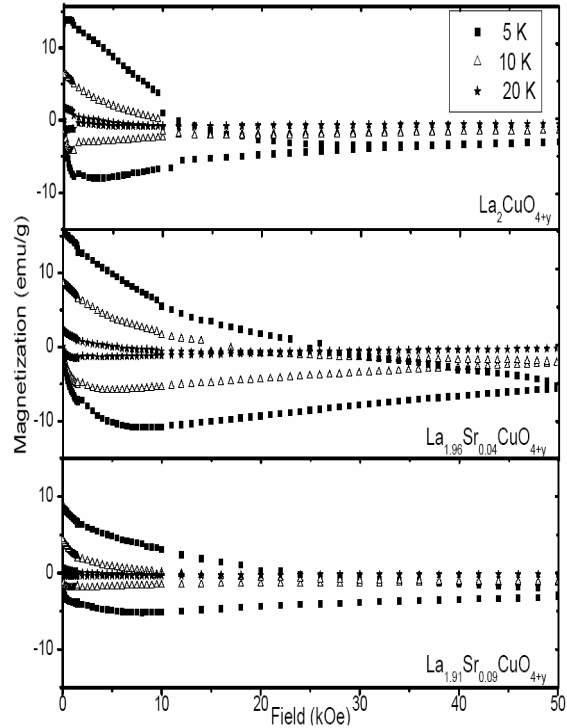

*FIG. 3 Dependence of magnetization on magnetic field (increasing and then decreasing) for various temperatures. The applied magnetic field is perpendicular to the c axis. Reversibility of magnetization at large fields are lower than what is observed for the field parallel to the c axis.*

$H_{irr}$ is not well determined for the $La_{2-x}Sr_xCuO_4$ samples but appears to be five

times or greater than for the super-oxygenated samples. We conclude that there is substantially less flux pinning in the super-oxygenated samples as compared to most high temperature superconductors with similar overall hole concentration.

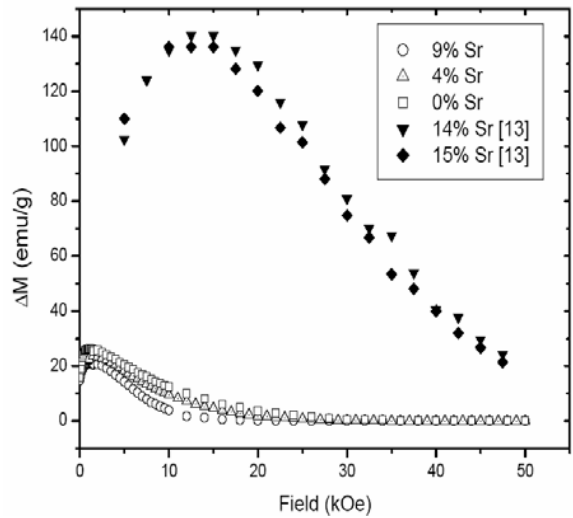

*FIG. 4. The graph of ΔM versus the magnetic field. The ΔM values for the two optimally doped samples (x = 0.14 and 0.15) were derived from the Fig.1 of Kodama et al [13]. The others were calculated from our field dependence data. All data sets were taken at 10 K.*

In general, pinning sites in a superconductor are formed by regions with a suppressed superconducting order parameter. However, the site has to be approximately the same size as the superconducting correlation length. For 214-type superconductors, the in-plane correlation length is known to be about 10 Å [28]. Regions much larger than this which have a suppressed superconducting order parameter will tend to form a separate non-superconducting phase, and thus, not strongly interact with the superconductor.

There are some excellent studies of flux trapping using well controlled, artificially created defects that form regions with reduced superconductivity. These have typically involved either ion beam damage or impurity precipitates. For an example, enhanced flux pinning was observed in $Bi_2Sr_2CaCu_2O_{8+y}$ single crystals embedded with MgO particles by Zhao et al. [16]. In these experiments nanometer-sized particles of MgO were introduced into the superconductor crystals and the critical current density, $J_C$, was measured. $J_C$ is proportional to ΔM in our measurements. It was found that for MgO concentrations below 10 %, $J_C$ at low temperatures were increased by over a factor of three compared with the samples without MgO. However, for concentrations of MgO above 10 % $J_C$ begins to decrease with further increase of impurities. Zhao et at. further found that when the defect concentration is large, particles assemble to form regions that have sizes greater than 10 μm [16]. Apparently, these regions do not act as effective pinning sites.

Other experiments have reached similar conclusions. For an example, Rudnev et al. report an enhancement of the critical current density in $Bi_2Sr_2Ca_2Cu_3O_{10+\delta}$ when nano-dimensional inclusions of tantalum carbide, niobium carbide or niobium nitride are introduced [17]. When the sizes of the introduced regions exceed 30 nm the critical current density decreases. The larger pinning sites not only do not favor flux pinning but in fact reduce the overall trapped flux in a material. Pu et al, also studied a series of $Bi_{1.8}Pb_{0.4}Sr_2Ca_{2.2-x}Pr_xCu_3O_y$ samples with different amounts of Pr substitution [18]. When the concentration of Pr is low they act as normal-like defects and enhance flux pinning. However when the Pr doping is increased up to a certain level, Pr or the defects induced by the Pr doping will congregate into large defect regions and reduce flux pinning. Based on the pinning force scaling analysis and microstructure observations this work concludes that only *point-like* Pr defects significantly enhance flux trapping while large defect conglomerations reduce the total pinning. Thus there is a large amount of evidence that the total amount of pinning in a high $T_C$ superconducting sample depends on the number of point defects present only up to a value above which the defects are no longer small and isolated. When defects can conglomerate into larger structures that substantially exceed the size of the vortex cores then both the pinning strength and the total trapped flux decrease.

In light of these comparisons, our data not only reveal information concerning the size of the phase separated magnetic regions in our super-oxygenated samples, but also give an indication of how those regions evolve from the more commonly studied cation doped $La_{2-x}Sr_xCuO_4$. There are a variety of experiments, notably Scanning Tunneling Microscopy (STM) imaging, Nuclear Magnetic Resonance (NMR) and μSR studies, that indicate the presence of inhomogeneity on a short, nanometer scale in most cuprate superconductors [29 - 32]. On the other hand, there are several results that indicate that the separate magnetic and superconducting regions in the super-oxygenated samples are quite large. Among these are the presence of a Meissner signal itself, requiring superconducting regions on the order of a penetration depth. We have carried out high field μSR that reflect the behavior of a muon in a vortex lattice state [33]. The associated relaxation parameter matches that from optimally doped $La_{2-x}Sr_xCuO_4$ thus indicating that the penetration depth in both samples is about the same and that the superconducting regions in the super-oxygenated samples are at least as large as a penetration depth [33, 34]. Neutron scattering measurements of the magnetic phase by both our own group [35] and Lee et al. [36] show narrow peaks indicating an in plane coherence of 400 Å or greater.

Thus it is natural to consider that the primary difference between the phase separated, super-oxygenated compounds we study here and cation doped $La_{2-x}Sr_xCuO_4$ is the length scale associated with inhomogeneities, with the former having much larger phase separated regions. The nano-scale inhomogeneity reported for $La_{2-x}Sr_xCuO_4$ apparently provides many strong flux pinning centers. However, the much larger phase separated regions present in the super-oxygenated materials are not truly part of the superconducting matrix and not effective at pinning flux. A similar argument might also explain why there is substantial amount of flux pinning when the field is applied perpendicular to the c axis. In this case the flux is pinned between planes. Through an analysis of neutron peaks, Lee et al. [36], have reported that the magnetic regions extend for only 13 Å in the c direction. The shorter length scale perpendicular to the planes is more suitable for flux pinning and enhances the pinning energy. However, there are fewer pinning sites between the planes and as a result the over all flux trapping is small when compared to the c parallel case.

Taken together, the flux pinning results indicate that in some sense the separated superconducting regions in the super-oxygenated compounds are more pure than typical superconducting $La_{2-x}Sr_xCuO_4$; within the superconducting regions there are fewer areas of weak superconductivity to trap flux. There is less charge inhomogeneity in the superconducting regions of the fully phase separated compound than in the usual $La_{2-x}Sr_xCuO_4$. This might also explain why the super-oxygenated samples have critical temperatures above 40 K while the highest $T_C$ in cation doped $La_{2-x}Sr_xCuO_4$ is near 38 K. While charge inhomogeneity is beginning to appear ubiquitous in high temperature superconductors, it may not be essential to the phenomenon.

## V. Conclusion

The superconducting phase of super-oxygenated and phase separated $La_{2-x}Sr_xCuO_{4+y}$ traps little flux compared to most high temperature superconducting samples. The lack of flux trapping appears to be due to the separated magnetic regions being too large to act as pinning centers. Furthermore, the presence of relatively strong trapping in optimally cation doped $La_{2-x}Sr_xCuO_4$ without excess oxygen may be due to nanoscale inhomogeneities that act as flux trappers. This view indicates that most high temperature superconductors are inhomogeneous. Doping the structure with highly mobile interstitial oxygen allows for the inhomogeneous regions to grow to micron like length scales so that the separate regions are easily measured. The superconducting regions of these phase separated samples may provide a new materials system for exploring the superconducting properties of a more pure high Tc superconductor.

**References:**

1.  G. Bednorz and K. A. Muller, Z. Phys. B **64**, 189 (1986)

2. S. Etemad, D. E. Aspnes, M. K. Kelly, R. Thompson, J. –M. Tarascon, and G. W. Hull Phys. Rev. B **37**, 3396 (1988)

3.  B. O. Wells et al., Z. Phys. B **100**, 535 (1996 )

4. J. D. Jorgensen, B. Dabrowski, Shiyou Pei, D. G. Hinks, L. Soderholm, B. Morosin, J. E. Schirber, E. L. Venturini and D. S. Ginley, Phys. Rev. B. **38**, 11337 (1988)

5. C. Rial, E. Mor´an, , M. A. Alario-Franco, U. Amado, and N. H. Anderson, Physica *C* **254,** 233–248 (1995).

6. B. O. Wells et al., Science Vol. **277,** 1067 (1997)

7. A. R. Moodenbaugh, Youwen Xu, M. Suenaga, T. J. Folkerts, and R. N. Shelton, Phys. Rev. B **38,** 4596 (1988).

8. B. Khaykovich et al., Phys. Rev. B **66,** 014528 (2002)

9. A.T. Savici et al., Phys. Rev. B **66,** 014524 (2002)

10. H. E. Mohottala et al., Nature Materials **5**, 377 (2006)

11. K. Yamada, C. H. Lee, K. Kurahashi, J. Wada, S. Wakimoto, S. Ueki, H. Kimura, and Y. Endoh, Phys. Rev. B **57**, 6165 (1998)

12. Patrick A. Lee, Naota Nagaosa, and Xia-Geng Wen Rev. Mod. Phys. **78**, 17 (2006)

13. K. Lefmann et al., Journal of Law Temperature Physics, Vol. **135,** Nos. 5/6 621 (2004)

14**.** Y. Kodama, K. Oka, Y. Yamaguchi, Y. Nishihara and K. Kajimura Phys Rev. B **56,** 6265 (1997)

15. T. Kimura, K. Kashio, T. Kobayashi, Y. Nakayama, N.Motohira, K. Kitazawa, and Y. Yamafuji Physica C **192,** 247 (1992)

16. B. Zhao, W. H. Song, X. C. Wu, W. D. Huang, M. H. Pu, J. J. Du, Y. P. Sun and H. C. Ku Supercond. Sci. Technol. **13**, 165 (2000)

17. I. A. Rudnev, B. P. Mikhailov, and P. V. Bobin Technical Physics Letters, Vol. **31,** No. 2, 176 (2005)

18. M. H. Pu, W H Song, B, X C Wu, T Hu, Y P Sun and J. J. Du Supercond. Sci. Technol. **14,** 305 (2001)

19. I. M. Obaidat, and B. A. Albiss, Supercond. Sci. Technol. **19,** 151 (2006)

20. J. W. Rogers Jr., N.D. Shinn, J. E. Schirber, E. L. Venturini , D.S. Ginley, and B. Morosin, Phys. Rev. B. **38,** 5021 (1988)

21. I. Tanaka and H. Kojima, Nature **337,** 21 (1989)

22. F. C Chou, N. R Belk, M. A. Kastner, R. J. Birgeneau, and A. Aharony, Phys. Rev. Lett. **75**, 2204 (1995).

23. D. C. Johnston, et al. Physica *C* **235-240,** 257 (1994)

24. Charles P. Bean, Rev. Mod. Phys. **36**, 31 (1964)

25. T. R. Dinger, T. K. Worthington, W. J. Gallagher, and R. L. Sandstrom, Phys. Rev. Lett. **58**, 2687 (1987)

26. K. Kishio et al., Supercond. Sci. Technol. **5,** S69 (1992)

27. F. C. Chou, D. C. Johnston, S-W. Cheong, and P. C. Canfield, Physica C **216**, 66 (1993)

28. V. J. Emery and S. A. Kivelson "Recent Developments in High Temperature Superconductivity" – Publisher, Springer Berlin / Heidelberg Vol. **475**, pg 9-13 (1996)

29. K. M. Lang, V. Madhavan, J. E. Hoffman, E. W. Hudson, H. Eisaki, S. Uchida, and J. C. Davis, *Nature* **415**, 412, (2002).

30. P. M Singer, A. W. Hunt, and T. Imai, Phys. Rev. Lett. **88**, 047602-(1-4), (2002).

31. Ch. Niedermayer, C. Bernhard, T. Blasius, A. Golnik, A. Moodenbaugh, and J. I. Budnick, Phys. Rev. Lett. **80**, 3843,(1998)

32. C. Panagopoulos , B.D. Rainford , J. R. Cooper , and C.A. Scott, Physica C **341-348**, 843 (2000)

33. Mohottala et al - to be published

34. E. J. Ansaldo, J. H. Brewer, T. M. Riseman, J. E. Schirber, E. L. Venturini, B. Morosin, D. S. Ginley, and B. Sternlieb, Phys. Rev. B **40**, 2555, (1989)

35. Udby et al – to be published

36. Y. S. Lee et al., Phys. Rev. B **60,** 3643 (1999)